\documentclass[aps,pra,preprint,groupedaddress]{revtex4-1}
\usepackage{graphicx}

\begin{document}


\title{Enhanced principle component method for fringe removal in cold atom images}



\author{Feng Xiong}
\affiliation{
School of Physics, Georgia Institute of Technology\\
837 State Street NW\\
Atlanta, GA, 30332
}

\author{Yun Long}
\affiliation{
School of Physics, Georgia Institute of Technology\\
837 State Street NW\\
Atlanta, GA, 30332
}

\author{Colin V. Parker}
\email{cparker@gatech.edu}
\affiliation{
School of Physics, Georgia Institute of Technology\\
837 State Street NW\\
Atlanta, GA, 30332
}


\date{\today}

\begin{abstract}
Many powerful imaging techniques for cold atoms are based on determining the optical density by comparing a beam image having passed through the atom cloud to a reference image taken under similar conditions with no atoms. In practice the beam profile typically contains interference fringes whose phase is not stable between camera exposures. To reduce the error of these fringes in the computed optical density, an algorithm based on principle component analysis (PCA) is often employed. However, PCA is general purpose and not tailored to the specific case of interference fringes. Here we demonstrate an algorithm that takes advantage of the Fourier-space structure of interference fringes to further reduce the residual fringe signatures in the optical density.
\end{abstract}


\maketitle
\tableofcontents
\section{Introduction}
Laser-cooled cold and ultracold atoms serve a variety of purposes in quantum simulation, quantum sensing, and optical spectroscopy. For many of these, detailed measurements of the spatial distribution of the atoms are required. Many methods exist to determine this distribution, with two popular methods being absorption imaging and phase contrast imaging. Absorption imaging is a general purpose technique which passes a probe beam through the atomic cloud, leaving a shadow in the transmitted beam due to atomic absorption, from which the name derives. Although the signal is not background-free, as it would be for a fluorescence or a bright-field image, this technique is nonetheless powerful and can be used even in situations with small atom number such as on atom chips \cite{smith_absorption_2011}, or where only a few atoms are present \cite{streed_absorption_2012,konstantinidis_atom_2012}. This comes from the ability to pass a reference beam through the sample without atoms present (or with the laser off resonance), and the optical density can be determined from the ratio of the signals with and without atoms. Phase contrast imaging involves a similar setup, but the probe beam is first brought to a focus, at which point the beam center is phase shifted by a small piece of material of the correct thickness\cite{andrews_propagation_1997,higbie_direct_2005,vengalattore_periodic_2010}. Acting on the beam center at the focus affects primarily the portion of the beam which was unaffected by the atoms, so that the atom-influenced portion of the light is overall phase shifted relative to the background, and rather than absorption one detects the phase change induced by refractive shifts. In other words, phase advancement or retardation is changed into reduction or enhancement of the background by the phase shift. This can be useful for the detection of atomic spin states, for example.

Both of the techniques require as the final step the comparison of local intensity ratios between two images. Unfortunately, probe beams must be spectrally narrow and have long coherence lengths, which leads to interference fringes from pairs of parallel surfaces in the imaging system. As it is often convenient to propagate the imaging beam through many optical elements, including waveplates, beamsplitters, lenses, and vacuum chamber windows, it is impractical to eliminate all of these fringes. Furthermore because the apparatus often has sources of vibration or temperature drift, the fringes will typically change between subsequent exposures, so that even nominally identical reference exposures will be different. Fortunately, it often possible to correct for many of these changes through post-processing, for example by using principle component analysis (PCA)\cite{li_reduction_2007,segal_revealing_2010,cao_extraction_2019}.

In this work we introduce a novel enhancement of the PCA method, based on the Fourier-space patterns typical of interference fringes. The layout of the paper is as follows: first we introduce the basic method of principle component analysis as applied to cold atom imaging. Then, we discuss our enhanced method, and show comparisons based on example data from a real cold atom apparatus\cite{long_all-optical_2018}. Finally, we offer some concluding remarks and discuss possible further refinements.

\section{Basic Principle Component Analysis for Fringe Removal}
In this section we outline the traditional approach to fringe removal using PCA. We begin by assuming a sequence of $n$ cold atom images, each consisting of the probe beam, or an ``atomic'' exposure, containing a region of lower or higher optical intensity induced by the atoms, and a reference beam, or ``light'' exposure. When dark current is an issue in the imaging system, both of these can have a ``dark'' exposure subtracted. Note also that for purposes of the algorithm it is not strictly necessary that the number of atom and light images be equal. We treat the dark-subtracted images as vectors (i.e. by indexing them column first), and form the image sequences into matrices $\mathbf{A}_{ij}$ and $\mathbf{L}_{ij}$, respectively, where $i$ indexes the pixel and $1 \le j \le n$ indexes the image. In what follows we will drop the subscripts. To compute the optical density, we wish to find the hypothetical matrix $\mathbf{A}'$ that represents what the imaging beam would have looked like in each exposure if the atoms had not been present. Then the optical density can be computed as
\begin{equation}
    \mathbf{OD} = log(\mathbf{A}) - log(\mathbf{A}'),
    \label{eq:ODeqn}
\end{equation}
where the logarithm is taken element-wise. One approach to guess $\mathbf{A}'$ is to approximate $\mathbf{A}' \approx \mathbf{L}$, but this is best suited to configurations where pairs of successive images are very similar, and has potentially large error if the interference fringes differ between such pairs. Another simple method is to compute the ``mean image'' matrix $\mathbf{\tilde{L}}$, obtained by replacing every row (corresponding to a particular image pixel) of $\mathbf{L}$ with the mean value of that row, and approximate $\mathbf{A}' \approx \mathbf{\tilde{L}}$. We call this ``no fringe removal'', and it suffers from a similar problem. However, it is useful as a first step, and in what follows we subtract the mean image to work explicitly with differences from the mean, forming the matrices $\delta\mathbf{A} = \mathbf{A}-\mathbf{\tilde{L}}$ and $\delta\mathbf{L} = \mathbf{L}-\mathbf{\tilde{L}}$ while attempting reconstruct $\mathbf{A}' = \delta\mathbf{A}'+\mathbf{\tilde{L}}$.

A more sophisticated approach is possible if we can rely on having a region of the imaging beam that is known to contain no atomic signal. Denoting the matrix that projects into the atom-free subspace with $\mathbf{P}$, we seek to use the information in $\mathbf{P}\delta \mathbf{A}$ in order to guess $\delta \mathbf{A}'$. That is, we are trying to use the information at the edge of the beam to guess what the middle of the beam should have looked like with no atoms. If we have a ``fringe matrix'' $\mathbf{F}$ consisting of a sequence of images of fringes, we can attempt to construct $\delta \mathbf{A}'$ using the coefficient matrix $\mathbf{C}$ from a least-squares fit,
\begin{equation}
    \mathbf{P}\mathbf{F}\mathbf{C} = \mathbf{P}\delta \mathbf{A},
    \label{eq:leastsquaresfit}
\end{equation}
to reconstruct the full matrix $\delta \mathbf{A}' \approx \mathbf{F}\mathbf{C}$, which becomes
\begin{equation}
    \delta \mathbf{A}' \approx \mathbf{F}(\mathbf{F}^T\mathbf{P}\mathbf{F})^{-1}\mathbf{F}^T\mathbf{P}\delta\mathbf{A}.
    \label{eq:leastsquaressoln}
\end{equation}

So far we should note that the method is completely general and applicable to any approach to estimating $\delta \mathbf{A}'$, including our novel approach. The differences lie in how to generate the fringe matrix $\mathbf{F}$. The simplest method is to choose the light image matrix $\delta \mathbf{L}$. This is a subset of a more general PCA approach, in which the covariance matrix $\mathbf{K} = (\delta \mathbf{L}^T)\delta\mathbf{L}$ is diagonalized and the eigenvectors corresponding to some chosen number of the largest eigenvalues are used to form $\mathbf{F}$. For typical cold atom images, this method has a few limitations. There are typically a very large number of possible sources of fringes (each pair of parallel surfaces can yield a set), and there may not be enough independent observations to account for every variation of every fringe, particularly in smaller data sets. Furthermore, to completely fit a sinusoidal fringe in a linear approach requires a two-dimensional basis per fringe (for the sine and cosine component), also a challenge with a small number of images. If the fringe phase shifts by a few tens of degrees during light images but systematically moves by 90 degrees between the atomic image and the light image, the atomic image may be too far out of sample to be accurately fit. Conversely, keeping large inventories of thousands of background images means that minor modifications to the imaging beam path (whether intentional or not) require a new inventory to be acquired.

\section{Enhancing the fringe data set}
\begin{figure}
\includegraphics[width=0.5\textwidth]{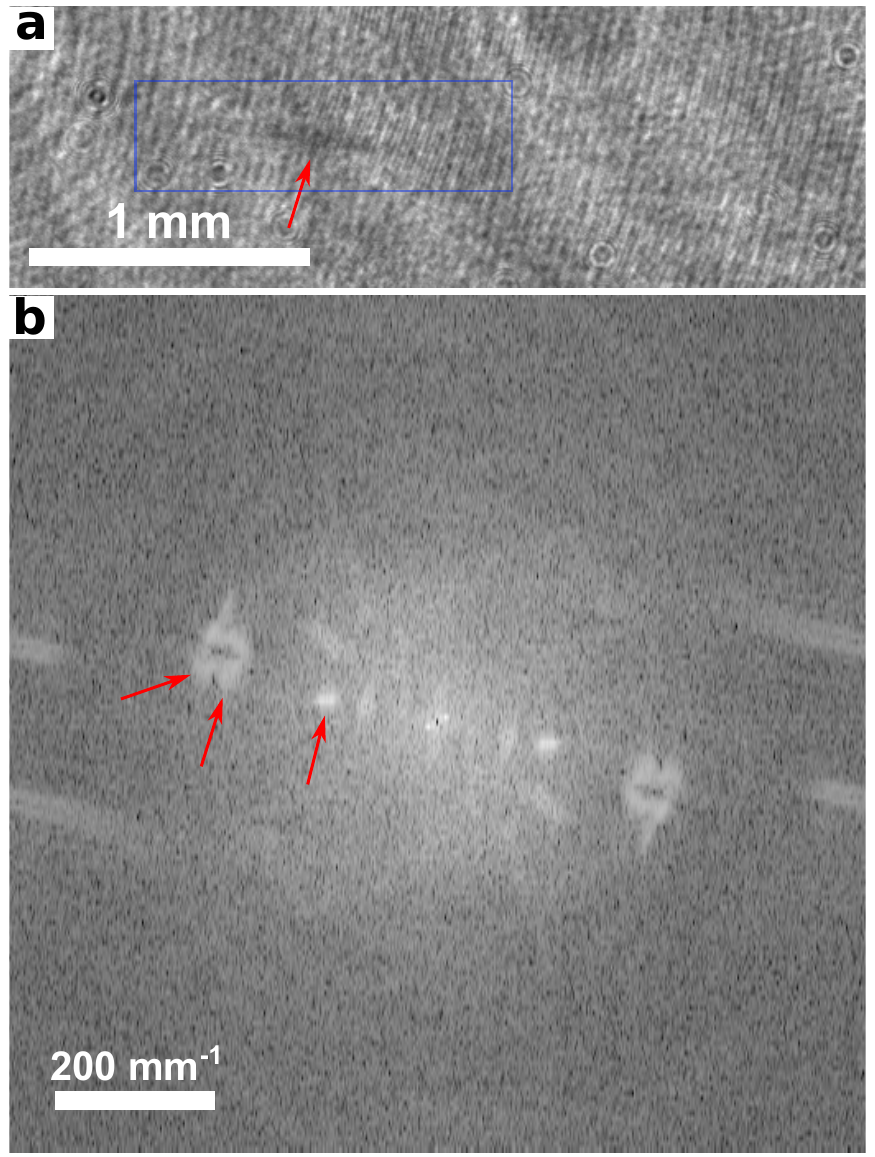}
\caption{\label{fig:fringe_patterns} {\bf a}. A typical absorption image from our apparatus. The arrow indicates the atomic absorption profile, the box indicates the atomic region that is excluded when fitting the background. {\bf b}. The DFT of this image. Arrows indicate the distinct peaks corresponding to interference fringes.}
\end{figure}

Our enhancement consists of the ability to ``guess'' more complete fringe patterns based on understanding their structure in Fourier space. Typically, interference fringes show up as sets of concentric rings, whose structure in Fourier space would be similarly ring-like. However, the portion of the field of view containing the atoms does not contain the centers for most rings, so that many of these patterns are nearly plane waves over the field of view, and hence correspond with distinct peaks in Fourier space. Figure \ref{fig:fringe_patterns} shows a typical beam image from our apparatus, along with it's discrete Fourier transform (DFT).

\begin{figure}
\includegraphics[width=\linewidth]{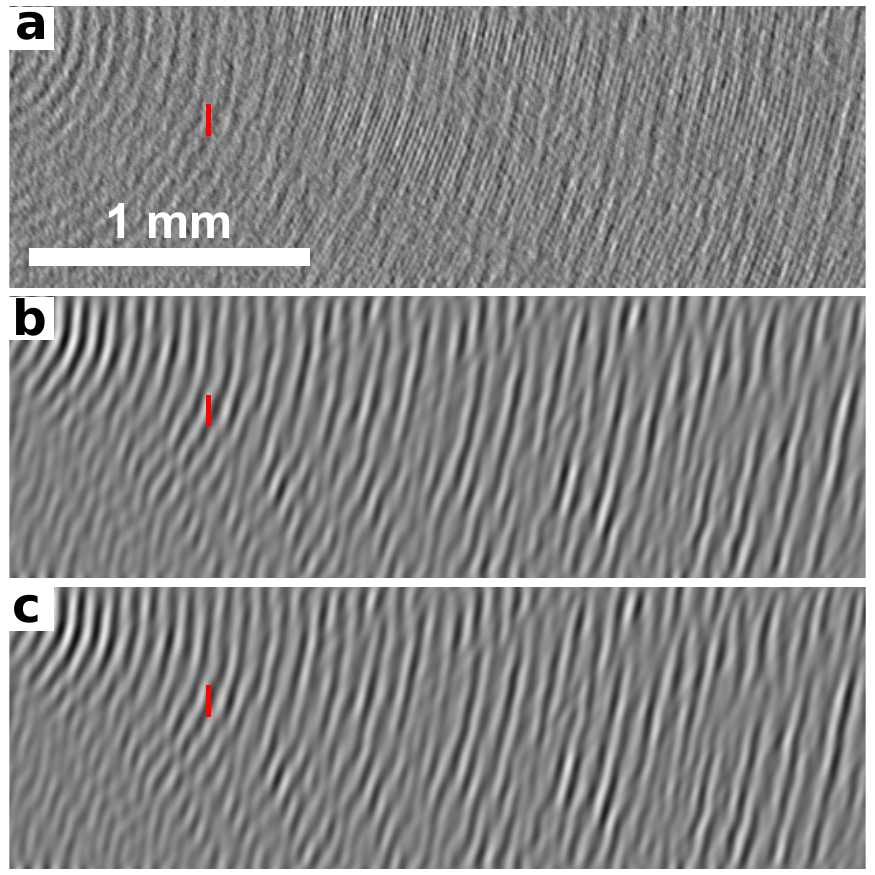}
\caption{{\bf a}. The largest principle component from a set of 54 reference images. {\bf b}. The real part of the filtered fringe profile. {\bf c}. The imaginary part of the filtered fringe profile. The bars are at a fixed position to show the phase shift. Note that atoms are present in near the center of this image, which generate interference patterns of their own at this contrast level.}
\label{fig:fringe_fits}
\end{figure}

The presence of peaks in the DFT motivates our modified PCA approach. The algorithm works as follows. First, as in conventional PCA, we determine the largest eigenvector of the covariance matrix of the reference images. Figure \ref{fig:fringe_fits}a shows the corresponding image in our reference data set.

Next, we compute the DFT and identify the largest Fourier component. This is done simply by selecting one of the pixels with the highest amplitude (since the original image is real most amplitudes will occur in pairs). Since eventually many components will be considered, it is not detrimental that noise may influence this selection. We then apply a Gaussian filter to the DFT centered on this pixel, and perform the inverse DFT. This provides a two-dimensional bandpass filter with a moderate length scale (corresponding to $30 \textrm{ }\mu\textrm{m}$ in the images shown). Importantly, the resulting filtered image is no longer purely real, but instead contains an imaginary part corresponding to a phase shifted version of the fringes. The filtered real and imaginary parts of the first principle component are shown in Fig. \ref{fig:fringe_fits}b-c. Note that the filtering process is non-linear because the center is chosen based on the highest amplitude Fourier component.

Our algorithm then proceeds as follows. We fit the filtered fringe pattern to the dataset and re-compute the principle component of the residual, apply the filter again, and produce a second filtered fringe pattern. This process is repeated, each time fitting using the entire accumulated set of fringe patterns until the filtered fringe pattern set is as large as desired. Because the filtering process is non-linear, there is in principle no limit to the number of fringe patterns than can be generated from a small data set, other than that set by the number of pixels in the atom-excluded region of the image. This is in contrast to a linear approach, which could generate at most $N-1$ linearly independent fringe patterns from a set of $N$ independent images, after accounting for the subtraction of the mean.

We make two further refinements to this process to address some remaining observations. First, not all image variation follows a clear fringe structure, and second, as the fringe set grows larger, some of the information can become redundant. To address the first issue, we periodically add unfiltered principle components. We have found that adding 1 unfiltered principle component per 8 filtered fringe patterns is a good compromise. To address the second issue, we periodically reduce the set of fringe patterns by performing a principle component analysis in the space of the fringe patterns. That is, given a partial fringe set $\mathbf{F}'$ we perform a least squares fit for the coefficients $\mathbf{C}'$ in
\begin{equation}
    \mathbf{F}'\mathbf{C}' = \delta \mathbf{L},
\end{equation}
and extract the largest eigenvectors from the matrix $\mathbf{C}'^{T}\mathbf{C}'$. At this point the fits are done over the complex numbers, which gives an estimate for which linear combinations of fringe patterns are significant and should be kept, but is not optimal for actual fitting of a real-valued image.
Our complete algorithm involves multiple rounds, each corresponding to adding 8 filtered principle components plus one unfiltered component, and then then removing the 4 least significant components from the entire data set. Running this process 15 times therefore yields a set of 75 fringe patterns, each of which is complex. To establish the final set of fringe patterns ($\mathbf{F}$), we take the real and imaginary part to produce a linearly independent set of 150 patterns.

\section{Comparison of methods}
\begin{figure}
\includegraphics[width=\linewidth]{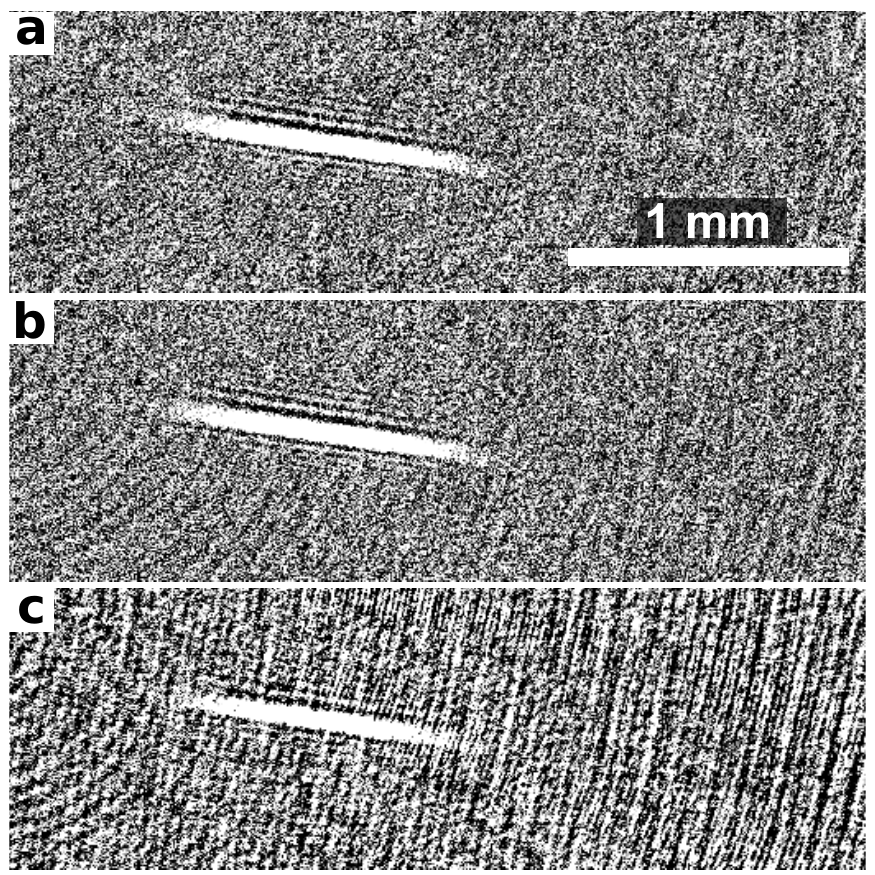}
\caption{Optical density images for the same image generated with the enhanced ({\bf a}) and simple ({\bf b}) PCA fringe removal, and with no fringe removal at all ({\bf c}). }
\label{fig:fringe_compare}
\end{figure}

Figure \ref{fig:fringe_compare} shows a comparison of the enhanced PCA method and a simple PCA method.

Though both methods lead to significant reduction of the fringes, it can be seen that the enhanced method provides further reduction, which is particularly noticeable near the bottom of the image. To perform a more quantitative representation, we analyzed the variance of a set of images containing no atoms, produced by dividing up the reference image set into a ``training'' set and an ``analysis'' set. The first $n$ images were used for training, (i.e. taken as the reference images) and the final $n$ images for analysis (i.e. taken as the probe or atom images despite containing no atoms). This was done with values of $n$ equal to $5$, $10$, and $27$ (the largest possible with an initial set of 54 images). The variance is computed from the fit residual $\mathbf{A}-\mathbf{A}'$ only over the region where the atoms would be, and for which the data is unavailable to either algorithm.

\begin{figure}
\includegraphics[width=0.5\textwidth]{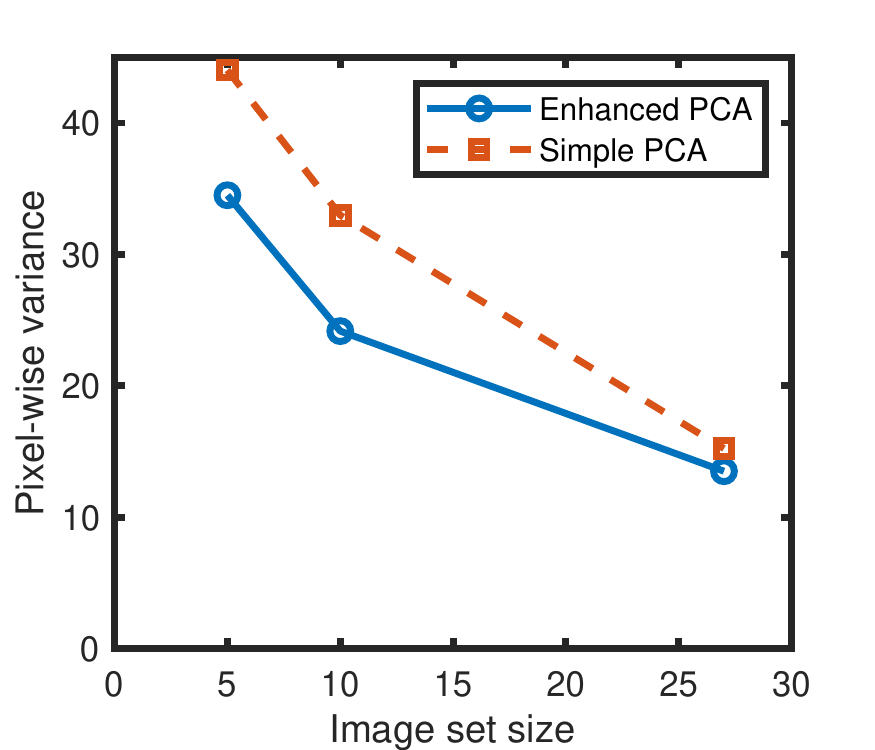}
\caption{\label{fig:fringe_plots}Comparison of the enhanced versus simple PCA methods on atom-free images as a function of sample image set size.}
\end{figure}

Figure \ref{fig:fringe_plots} shows the resulting variance for different image set sizes. For both methods, including a larger sample size results in lower variance, meaning more of the fringes have been removed. However, the enhanced method consistently beats the simple one. This is particularly true for smaller sample sets, which may be for two reasons. First, one of the strengths of the enhanced method is the ability to generate a larger linearly independent set of fringe patterns, which is most important for small data sets. Second, as more fringes are removed, there will be saturation to the intrinsic camera noise level. Since we have no method to completely eliminate fringes at present, we cannot estimate this noise level precisely, but using the square root of the number of raw CCD counts yields an estimate of 11, consistent with the observed saturation. In our cold atom apparatus, we believe that camera noise is the limiting factor in our atomic spin detection after implementing this method, which currently allows us to determine the spin population difference of a $^{6}\textrm{Li}$ gas with a statistical uncertainty of $6000$ atoms per shot.

\section{Conclusion}
We have presented a novel, enhanced PCA method for removing interference fringes from cold atom images. Compared with simple PCA, our method produces noticeably improved images, which can be quantified as a 25\% decrease in the pixel-wise variance for small data sets, and a 10\% decrease for larger ones. The improvement stems from recognizing the fringe structure as peaks in Fourier space (corresponding with plane waves in real space), which allows the construction of a linearly independent basis of high dimension even from a limited set of reference images. In the case of our imaging system, the enhanced method leaves the extracted optical density profiles near the shot-noise limit. However, in situations with higher photon counts the method could be extended to use any approximate method of fringe identification going beyond plane waves. For example, specific patterns of circular fringes, such as Airy disks, could be directly fit. Finally, we mention the possibility that reference images could be acquired in which the variation in the fringes is deliberately enhanced, which could allow the enhanced PCA algorithm to identify the fringe patterns more clearly.

\begin{acknowledgments}
We acknowledge support from the Air Force Office of Scientific Research, Young Investigator Program, through grant number FA9550-18-1-0047.
\end{acknowledgments}

\bibliography{references}

\begin{thebibliography}{10}%
\makeatletter
\providecommand \@ifxundefined [1]{%
 \@ifx{#1\undefined}
}%
\providecommand \@ifnum [1]{%
 \ifnum #1\expandafter \@firstoftwo
 \else \expandafter \@secondoftwo
 \fi
}%
\providecommand \@ifx [1]{%
 \ifx #1\expandafter \@firstoftwo
 \else \expandafter \@secondoftwo
 \fi
}%
\providecommand \natexlab [1]{#1}%
\providecommand \enquote  [1]{``#1''}%
\providecommand \bibnamefont  [1]{#1}%
\providecommand \bibfnamefont [1]{#1}%
\providecommand \citenamefont [1]{#1}%
\providecommand \href@noop [0]{\@secondoftwo}%
\providecommand \href [0]{\begingroup \@sanitize@url \@href}%
\providecommand \@href[1]{\@@startlink{#1}\@@href}%
\providecommand \@@href[1]{\endgroup#1\@@endlink}%
\providecommand \@sanitize@url [0]{\catcode `\\12\catcode `\$12\catcode
  `\&12\catcode `\#12\catcode `\^12\catcode `\_12\catcode `\%12\relax}%
\providecommand \@@startlink[1]{}%
\providecommand \@@endlink[0]{}%
\providecommand \url  [0]{\begingroup\@sanitize@url \@url }%
\providecommand \@url [1]{\endgroup\@href {#1}{\urlprefix }}%
\providecommand \urlprefix  [0]{URL }%
\providecommand \Eprint [0]{\href }%
\providecommand \doibase [0]{http://dx.doi.org/}%
\providecommand \selectlanguage [0]{\@gobble}%
\providecommand \bibinfo  [0]{\@secondoftwo}%
\providecommand \bibfield  [0]{\@secondoftwo}%
\providecommand \translation [1]{[#1]}%
\providecommand \BibitemOpen [0]{}%
\providecommand \bibitemStop [0]{}%
\providecommand \bibitemNoStop [0]{.\EOS\space}%
\providecommand \EOS [0]{\spacefactor3000\relax}%
\providecommand \BibitemShut  [1]{\csname bibitem#1\endcsname}%
\let\auto@bib@innerbib\@empty
\bibitem [{\citenamefont {Smith}\ \emph {et~al.}(2011)\citenamefont {Smith},
  \citenamefont {Aigner}, \citenamefont {Hofferberth}, \citenamefont {Gring},
  \citenamefont {Andersson}, \citenamefont {Wildermuth}, \citenamefont
  {Krüger}, \citenamefont {Schneider}, \citenamefont {Schumm},\ and\
  \citenamefont {Schmiedmayer}}]{smith_absorption_2011}%
  \BibitemOpen
  \bibfield  {author} {\bibinfo {author} {\bibfnamefont {D.~A.}\ \bibnamefont
  {Smith}}, \bibinfo {author} {\bibfnamefont {S.}~\bibnamefont {Aigner}},
  \bibinfo {author} {\bibfnamefont {S.}~\bibnamefont {Hofferberth}}, \bibinfo
  {author} {\bibfnamefont {M.}~\bibnamefont {Gring}}, \bibinfo {author}
  {\bibfnamefont {M.}~\bibnamefont {Andersson}}, \bibinfo {author}
  {\bibfnamefont {S.}~\bibnamefont {Wildermuth}}, \bibinfo {author}
  {\bibfnamefont {P.}~\bibnamefont {Krüger}}, \bibinfo {author} {\bibfnamefont
  {S.}~\bibnamefont {Schneider}}, \bibinfo {author} {\bibfnamefont
  {T.}~\bibnamefont {Schumm}}, \ and\ \bibinfo {author} {\bibfnamefont
  {J.}~\bibnamefont {Schmiedmayer}},\ }\href {\doibase 10.1364/OE.19.008471}
  {\bibfield  {journal} {\bibinfo  {journal} {Optics Express}\ }\textbf
  {\bibinfo {volume} {19}},\ \bibinfo {pages} {8471} (\bibinfo {year}
  {2011})}\BibitemShut {NoStop}%
\bibitem [{\citenamefont {Streed}\ \emph {et~al.}(2012)\citenamefont {Streed},
  \citenamefont {Jechow}, \citenamefont {Norton},\ and\ \citenamefont
  {Kielpinski}}]{streed_absorption_2012}%
  \BibitemOpen
  \bibfield  {author} {\bibinfo {author} {\bibfnamefont {E.~W.}\ \bibnamefont
  {Streed}}, \bibinfo {author} {\bibfnamefont {A.}~\bibnamefont {Jechow}},
  \bibinfo {author} {\bibfnamefont {B.~G.}\ \bibnamefont {Norton}}, \ and\
  \bibinfo {author} {\bibfnamefont {D.}~\bibnamefont {Kielpinski}},\ }\href
  {\doibase 10.1038/ncomms1944} {\bibfield  {journal} {\bibinfo  {journal}
  {Nature Communications}\ }\textbf {\bibinfo {volume} {3}},\ \bibinfo {pages}
  {933} (\bibinfo {year} {2012})}\BibitemShut {NoStop}%
\bibitem [{\citenamefont {Konstantinidis}\ \emph {et~al.}(2012)\citenamefont
  {Konstantinidis}, \citenamefont {Pappa}, \citenamefont {Wikström},
  \citenamefont {Condylis}, \citenamefont {Sahagun}, \citenamefont {Baker},
  \citenamefont {Morizot},\ and\ \citenamefont
  {Klitzing}}]{konstantinidis_atom_2012}%
  \BibitemOpen
  \bibfield  {author} {\bibinfo {author} {\bibfnamefont {G.}~\bibnamefont
  {Konstantinidis}}, \bibinfo {author} {\bibfnamefont {M.}~\bibnamefont
  {Pappa}}, \bibinfo {author} {\bibfnamefont {G.}~\bibnamefont {Wikström}},
  \bibinfo {author} {\bibfnamefont {P.}~\bibnamefont {Condylis}}, \bibinfo
  {author} {\bibfnamefont {D.}~\bibnamefont {Sahagun}}, \bibinfo {author}
  {\bibfnamefont {M.}~\bibnamefont {Baker}}, \bibinfo {author} {\bibfnamefont
  {O.}~\bibnamefont {Morizot}}, \ and\ \bibinfo {author} {\bibfnamefont
  {W.}~\bibnamefont {Klitzing}},\ }\href {\doibase 10.2478/s11534-012-0108-x}
  {\bibfield  {journal} {\bibinfo  {journal} {Open Physics}\ }\textbf {\bibinfo
  {volume} {10}} (\bibinfo {year} {2012}),\
  10.2478/s11534-012-0108-x}\BibitemShut {NoStop}%
\bibitem [{\citenamefont {Andrews}\ \emph {et~al.}(1997)\citenamefont
  {Andrews}, \citenamefont {Kurn}, \citenamefont {Miesner}, \citenamefont
  {Durfee}, \citenamefont {Townsend}, \citenamefont {Inouye},\ and\
  \citenamefont {Ketterle}}]{andrews_propagation_1997}%
  \BibitemOpen
  \bibfield  {author} {\bibinfo {author} {\bibfnamefont {M.~R.}\ \bibnamefont
  {Andrews}}, \bibinfo {author} {\bibfnamefont {D.~M.}\ \bibnamefont {Kurn}},
  \bibinfo {author} {\bibfnamefont {H.-J.}\ \bibnamefont {Miesner}}, \bibinfo
  {author} {\bibfnamefont {D.~S.}\ \bibnamefont {Durfee}}, \bibinfo {author}
  {\bibfnamefont {C.~G.}\ \bibnamefont {Townsend}}, \bibinfo {author}
  {\bibfnamefont {S.}~\bibnamefont {Inouye}}, \ and\ \bibinfo {author}
  {\bibfnamefont {W.}~\bibnamefont {Ketterle}},\ }\href {\doibase
  10.1103/PhysRevLett.79.553} {\bibfield  {journal} {\bibinfo  {journal}
  {Physical Review Letters}\ }\textbf {\bibinfo {volume} {79}},\ \bibinfo
  {pages} {553} (\bibinfo {year} {1997})}\BibitemShut {NoStop}%
\bibitem [{\citenamefont {Higbie}\ \emph {et~al.}(2005)\citenamefont {Higbie},
  \citenamefont {Sadler}, \citenamefont {Inouye}, \citenamefont {Chikkatur},
  \citenamefont {Leslie}, \citenamefont {Moore}, \citenamefont {Savalli},\ and\
  \citenamefont {Stamper-Kurn}}]{higbie_direct_2005}%
  \BibitemOpen
  \bibfield  {author} {\bibinfo {author} {\bibfnamefont {J.~M.}\ \bibnamefont
  {Higbie}}, \bibinfo {author} {\bibfnamefont {L.~E.}\ \bibnamefont {Sadler}},
  \bibinfo {author} {\bibfnamefont {S.}~\bibnamefont {Inouye}}, \bibinfo
  {author} {\bibfnamefont {A.~P.}\ \bibnamefont {Chikkatur}}, \bibinfo {author}
  {\bibfnamefont {S.~R.}\ \bibnamefont {Leslie}}, \bibinfo {author}
  {\bibfnamefont {K.~L.}\ \bibnamefont {Moore}}, \bibinfo {author}
  {\bibfnamefont {V.}~\bibnamefont {Savalli}}, \ and\ \bibinfo {author}
  {\bibfnamefont {D.~M.}\ \bibnamefont {Stamper-Kurn}},\ }\href {\doibase
  10.1103/PhysRevLett.95.050401} {\bibfield  {journal} {\bibinfo  {journal}
  {Physical Review Letters}\ }\textbf {\bibinfo {volume} {95}},\ \bibinfo
  {pages} {050401} (\bibinfo {year} {2005})}\BibitemShut {NoStop}%
\bibitem [{\citenamefont {Vengalattore}\ \emph {et~al.}(2010)\citenamefont
  {Vengalattore}, \citenamefont {Guzman}, \citenamefont {Leslie}, \citenamefont
  {Serwane},\ and\ \citenamefont {Stamper-Kurn}}]{vengalattore_periodic_2010}%
  \BibitemOpen
  \bibfield  {author} {\bibinfo {author} {\bibfnamefont {M.}~\bibnamefont
  {Vengalattore}}, \bibinfo {author} {\bibfnamefont {J.}~\bibnamefont
  {Guzman}}, \bibinfo {author} {\bibfnamefont {S.~R.}\ \bibnamefont {Leslie}},
  \bibinfo {author} {\bibfnamefont {F.}~\bibnamefont {Serwane}}, \ and\
  \bibinfo {author} {\bibfnamefont {D.~M.}\ \bibnamefont {Stamper-Kurn}},\
  }\href {\doibase 10.1103/PhysRevA.81.053612} {\bibfield  {journal} {\bibinfo
  {journal} {Physical Review A}\ }\textbf {\bibinfo {volume} {81}},\ \bibinfo
  {pages} {053612} (\bibinfo {year} {2010})}\BibitemShut {NoStop}%
\bibitem [{\citenamefont {Li}(2007)}]{li_reduction_2007}%
  \BibitemOpen
  \bibfield  {author} {\bibinfo {author} {\bibfnamefont {X.}~\bibnamefont
  {Li}},\ }\href@noop {} {\bibfield  {journal} {\bibinfo  {journal} {CHINESE
  OPTICS LETTERS}\ }\textbf {\bibinfo {volume} {5}},\ \bibinfo {pages} {3}
  (\bibinfo {year} {2007})}\BibitemShut {NoStop}%
\bibitem [{\citenamefont {Segal}\ \emph {et~al.}(2010)\citenamefont {Segal},
  \citenamefont {Diot}, \citenamefont {Cornell}, \citenamefont {Zozulya},\ and\
  \citenamefont {Anderson}}]{segal_revealing_2010}%
  \BibitemOpen
  \bibfield  {author} {\bibinfo {author} {\bibfnamefont {S.~R.}\ \bibnamefont
  {Segal}}, \bibinfo {author} {\bibfnamefont {Q.}~\bibnamefont {Diot}},
  \bibinfo {author} {\bibfnamefont {E.~A.}\ \bibnamefont {Cornell}}, \bibinfo
  {author} {\bibfnamefont {A.~A.}\ \bibnamefont {Zozulya}}, \ and\ \bibinfo
  {author} {\bibfnamefont {D.~Z.}\ \bibnamefont {Anderson}},\ }\href {\doibase
  10.1103/PhysRevA.81.053601} {\bibfield  {journal} {\bibinfo  {journal}
  {Physical Review A}\ }\textbf {\bibinfo {volume} {81}},\ \bibinfo {pages}
  {053601} (\bibinfo {year} {2010})}\BibitemShut {NoStop}%
\bibitem [{\citenamefont {Cao}\ \emph {et~al.}(2019)\citenamefont {Cao},
  \citenamefont {Tang}, \citenamefont {Guo}, \citenamefont {Chen},
  \citenamefont {Zhang},\ and\ \citenamefont {Zhou}}]{cao_extraction_2019}%
  \BibitemOpen
  \bibfield  {author} {\bibinfo {author} {\bibfnamefont {S.}~\bibnamefont
  {Cao}}, \bibinfo {author} {\bibfnamefont {P.}~\bibnamefont {Tang}}, \bibinfo
  {author} {\bibfnamefont {X.}~\bibnamefont {Guo}}, \bibinfo {author}
  {\bibfnamefont {X.}~\bibnamefont {Chen}}, \bibinfo {author} {\bibfnamefont
  {W.}~\bibnamefont {Zhang}}, \ and\ \bibinfo {author} {\bibfnamefont
  {X.}~\bibnamefont {Zhou}},\ }\href {\doibase 10.1364/OE.27.012710} {\bibfield
   {journal} {\bibinfo  {journal} {Optics Express}\ }\textbf {\bibinfo {volume}
  {27}},\ \bibinfo {pages} {12710} (\bibinfo {year} {2019})}\BibitemShut
  {NoStop}%
\bibitem [{\citenamefont {Long}\ \emph {et~al.}(2018)\citenamefont {Long},
  \citenamefont {Xiong}, \citenamefont {Gaire}, \citenamefont {Caligan},\ and\
  \citenamefont {Parker}}]{long_all-optical_2018}%
  \BibitemOpen
  \bibfield  {author} {\bibinfo {author} {\bibfnamefont {Y.}~\bibnamefont
  {Long}}, \bibinfo {author} {\bibfnamefont {F.}~\bibnamefont {Xiong}},
  \bibinfo {author} {\bibfnamefont {V.}~\bibnamefont {Gaire}}, \bibinfo
  {author} {\bibfnamefont {C.}~\bibnamefont {Caligan}}, \ and\ \bibinfo
  {author} {\bibfnamefont {C.~V.}\ \bibnamefont {Parker}},\ }\href {\doibase
  10.1103/PhysRevA.98.043626} {\bibfield  {journal} {\bibinfo  {journal}
  {Physical Review A}\ }\textbf {\bibinfo {volume} {98}},\ \bibinfo {pages}
  {043626} (\bibinfo {year} {2018})}\BibitemShut {NoStop}%
\end{thebibliography}%

\end{document}